\documentstyle[12pt]{article}
\newcommand{\doe}{This work was supported by the Director, Office of Energy
                  Research, Division of Nuclear Physics of the Office of High
                  Energy and Nuclear Physics of the U.S. Department of Energy
                  under Contract No. DE-AC03-76SF00098.}

\begin{document}

\begin{titlepage}

\begin{flushright}
     {\large LBL-34246}
  \end{flushright}
\vskip 2\baselineskip
\renewcommand{\thefootnote}{\fnsymbol{footnote}}
\setcounter{footnote}{0}
\begin{center}
\baselineskip=24pt
\mbox{}\\[5ex]
{\Large HIJING 1.0: A Monte Carlo Program for Parton and Particle Production
in High Energy Hadronic and Nuclear Collisions{\footnote{\doe}}}\\[5ex]
\baselineskip=18pt
{\large Miklos Gyulassy}\\[2ex]
{\em Physics Department, Columbia University, New York, NY 10027}\\[2ex]
{\large Xin-Nian Wang}\\[2ex]
{\em Nuclear Science Division, Mailstop 70A-3307,
        Lawrence Berkeley Laboratory}\\
{\em University of California, Berkeley, CA 94720}\\
        \mbox{}\\[3ex]
\today\\[5ex]
\end{center}

\begin{abstract}
\normalsize
\baselineskip=24pt
        Based on QCD-inspired models for multiple jets production, we
developed a Monte Carlo program to study jet and the associated particle
production in high energy $pp$, $pA$ and $AA$ collisions. The physics behind
the program which includes multiple minijet production, soft excitation,
nuclear shadowing of parton distribution functions and jet interaction
in dense matter is briefly discussed. A detailed description of the
program and instructions on how to use it are given.
\end{abstract}

\end{titlepage}

\newcommand{\lsim}
{\ \raisebox{2.75pt}{$<$}\hspace{-9.0pt}\raisebox{-2.75pt}{$\sim$}\ }
\newcommand{\gsim}
{\ \raisebox{2.75pt}{$>$}\hspace{-9.0pt}\raisebox{-2.75pt}{$\sim$}\ }

\baselineskip=18pt
\parindent=0.25in
\abovedisplayskip=24pt
\belowdisplayskip=24pt

\begin{center}
{\Large\bf PROGRAM SUMMARY}
\end{center}

\noindent{\em Title of program}: HIJING 1.0\\
\vspace{0pt}\\
{\em Catalogue number}:\\
\vspace{0pt}\\
{\em Program obtainable from}: xnwang@nsdssd.lbl.gov\\
\vspace{0pt}\\
{\em Computer for which the program is designed}: VAX, VAXstation,
SPARCstation and other computers with a FORTRAN 77 compiler
compiler\\
\vspace{0pt}\\
{\em Computer}: SPARCstation ELC; {\em Installation}: Nuclear Science Division,
Lawrence Berkeley Laboratory, USA\\
\vspace{0pt}\\
{\em Operating system}: SunOS 4.1.1\\
\vspace{0pt}\\
{\em Programming language used}: FORTRAN 77\\
\vspace{0pt}\\
{\em High speed storage required}: 90k word\\
\vspace{0pt}\\
{\em No. of bits  in a word}: 32\\
\vspace{0pt}\\
{\em Peripherals used}: terminal for input, terminal or printer for output\\
\vspace{0pt}\\
{\em No. of lines in combined program and test deck}: 6397 \\
\vspace{0pt}
\vspace{0pt}\\
{\em Nature of the physical problem}\\
In high-energy hadron and nuclear interactions, multiple minijet
production becomes more and more important. Especially in relativistic
heavy-ion collisions, minijets are expected to dominate transverse
energy production in the central rapidity region. Particle production
and correlation due to minijets must be investigated in order to
recognize new physics of quark-gluon plasma formation.
Due to the complication of soft interactions,
minijet production can only be incorporated in a pQCD
inspired model. The parameters in this model have to be tested first
against the wide range of data in $pp$ collisions. When extrapolating
to heavy-ion collisions, nuclear effects such as parton shadowing and
final state interactions have to be considered.\\
\vspace{0pt}\\
{\em Method of solution}\\
        Based on a pQCD-inspired model, multiple minijet production
is combined together with Lund-type model for soft interactions. Within
this model, triggering on large $P_T$ jet production automatically
biases toward enhanced minijet production. Binary approximation and Glauber
geometry for multiple interaction are used to simulate $pA$ and $AA$
collisions. A parametrized parton distribution function inside
a nucleus is used to take into account parton shadowing. Jet quenching
is modeled by an assumed energy loss $dE/dz$ of partons traversing
the produced dense matter. A simplest color configuration is assumed
for the multiple jet system and Lund jet fragmentation model is used
for the hadronization.\\
\vspace{0pt}\\
{\em Restrictions on the complexity of the problem}\\
        The program is only valid for collisions with c.m.
energy ($\sqrt{s}$) above 4 GeV/n. For central $Pb+Pb$ collisions,
some arrays have to be extended above $\sqrt{s}=10$ TeV/n.\\
\vspace{0pt}\\
{\em Typical running time}\\
The running time largely depends on the energy and the type of
collisions. For example (not including initialization):\\
\begin{tabbing}
$Pb+Pb$(central) AAAA\= $\sqrt{s}$=6.4 TeV/n \= $\sim$ events/min \kill
$pp$ \> $\sqrt{s}$=200 GeV \> $\sim$ 700 events/min.\\
$pp$ \> $\sqrt{s}$=1.8 TeV \> $\sim$ 250 events/min.\\
$Au+Au$(central) \> $\sqrt{s}$=200 GeV/n \> $\sim$ 1 event/min.\\
$Pb+Pb$(central) \> $\sqrt{s}$=6.4 TeV/n \> $\sim$ 1 event/10 min.
\end{tabbing}
\mbox{}\\
{\em Unusual features of the program}\\
        The random number generator used in the program is a
VAX VMS system subroutine RAN(NSEED). When compiled on a
SPARCstation, {\tt -xl} flag should be used. This function
is not portable. Therefore, one should supply a random
number generator to replace this function whenever a problem
is encountered.

\newpage

\begin{center}
{\Large\bf LONG WRITE-UP}
\end{center}

\section{Introduction}

        One of the goals of ultrarelativistic heavy ion experiments is
to study the quark-gluon substructure of nuclear matter and the
possibility of a phase transition from hadronic matter to quark-gluon
plasma (QGP)\cite{review} at extremely high energy densities.
Unlike heavy ion collisions
at the existing AGS/BNL and SPS/CERN energies, most of the physical
processes occurring at very early times in the violent collisions of
heavy nuclei at RHIC/BNL and the proposed LHC/CERN energies involve
hard or semihard parton scatterings\cite{kaja}
which will result in enormous amount of jet production and
can be described in terms of perturbative QCD (pQCD).

        The concept of jets and their association with
hard parton scatterings has been well established in hadronic
interactions and they have been proven to play a major role
in every aspect of $p\overline{p}$ collisions at CERN
$\mbox{Sp}\overline{\mbox{p}}\mbox{S}$ and Fermilab Tevatron
energies\cite{geist}.  Experimentally, jets are identified
as hadronic clusters whose transverse energy $E_T$ can be
reconstructed from the calorimetrical study\cite{ua2jet,alba88}
of the events. However, when the transverse energy of a jet becomes
smaller, $E_T<5$ GeV, it is increasingly difficult to
resolve it from the underlying background\cite{ua1minijet},
though theoretically, we would
expect that hard parton scatterings must continue to lower
transverse momentum. We usually refer to those as minijets
whose transverse energy are too low to be resolved
experimentally but the associated parton scattering
processes may still be calculable via pQCD. Assuming
independent production, it has been shown that
the multiple minijets production is important in
$p\bar{p}$ interactions to account for the increase
of total cross section\cite{gaisser} and the violation
of Koba-Nielsen-Olesen (KNO) scaling of the charged
multiplicity distributions\cite{sjostrand,wang91a}.

        In high energy heavy ion collisions, minijets
have been estimated\cite{kaja} to produce 50\% (80\%) of
the transverse energy in central heavy ion collisions at
RHIC (LHC) energies. While not resolvable as distinct jets,
they would lead to a wide variety of correlations, as in
$pp$ or $p\bar{p}$ collisions, among observables such as
multiplicity, transverse momentum, strangeness, and
fluctuations that compete with the expected signatures
of a QGP. Therefore, it is especially important to
calculate these background processes. Furthermore,
the calculation could also provide the initial
condition to address the issues of thermalization
and equilibration of a quark gluon plasma. In this
respect, the interactions of high $P_T$ jets inside
the dense medium is also interesting since the
variation of jet quenching phenomenon may serve
as one of the signatures of the QGP transition\cite{gyu89}.

        To provide a theoretical laboratory for studying
jets in high-energy nuclear interactions and testing the
proposed signatures such as jet quenching\cite{gyu89},
we have developed  a Monte Carlo model,
HIJING (heavy ion jet interaction generator)\cite{hijing},
which combines a QCD inspired model for jet production with the
Lund model\cite{lund} for jet fragmentation.
The formulation of HIJING was guided by the Lund FRITIOF\cite{fritiof}
and Dual Parton model\cite{dpm} for soft $A+B$ reactions at
intermediate energies ($\sqrt{s}\lsim 20$ GeV/nucleon) and
the successful implementation of pQCD processes in
PYTHIA\cite{sjostrand,pythia}
model for hadronic collisions. HIJING is designed
mainly to explore the range of possible initial conditions
that may occur in relativistic heavy ion collisions. To study
the nuclear effects, we also included nuclear shadowing\cite{shadow1}
of parton structure functions and a schematic model of final
state interaction of high $P_T$ jets in terms of an effective
energy loss parameter, $dE/dz$\cite{wang92a,gyu92}.
At $pp$ and $p\bar{p}$ level,
HIJING also made an important effort to address the interplay
between low $P_T$ nonperturbative physics and the hard pQCD
processes. This Monte Carlo model has been tested extensively
against data on $p+p(\bar{p})$ over a wide energy range,
$\sqrt{s}=50$-1800 GeV and $p+A$, $A+A$ collisions at moderate
energies $\sqrt{s}\leq 20$  GeV/n \cite{hijing,hijingpp}.
However, in this version of HIJING program, the space-time
development of final state interaction among produced
partons\cite{geiger} and hadrons was not considered.

        In this paper, we present a detailed description of
the Monte Carlo program together with a brief summary of
physical motivations. Since the program uses subroutines
of PYTHIA to generate the kinetic variables for each
hard scattering and the associated radiations, and JETSET
for string fragmentation, we refer readers to the original
publications\cite{pythia,jetset} for the description of these
programs. The physics involved in HIJING has been discussed
extensively\cite{hijing,hijingpp,wang92a}. This paper is
intended to be a documented
reference for the overall structure and detailed description
of the program.

        The organization of the paper is as the following.
In Section 2, we give a brief review of the QCD inspired model
for multiple jets production and soft interaction in
nucleon-nucleon collisions. The nuclear effects on jet
production and fragmentation are discussed in Section 3.
Section 4 will give a detailed description of the program.
Finally in Section 5 we will give instructions on how to
use the program and some simple examples are provided.

\section{Parton Production in $pp$ Collisions}

        The QCD inspired model is based on the assumption of
independent production of multiple minijets. It determines
the number of minijets per nucleon-nucleon collisions.
For each hard or semihard interaction the kinetic variables of
the scattered partons are determined by calling PYTHIA\cite{pythia}
subroutines. The  scheme for the accompanying soft interactions
is similar to FRITIOF model\cite{fritiof} with some difference
in the successive soft excitation of the leading quarks or
diquarks and $P_T$ transfer involved.  Since minijet production
is dominated by gluon scatterings, we assume that quark
scatterings only involve valence quarks and restrict the subsequent
hard processes to gluon-gluon scatterings. Simplification is
also made for the color flow in the case of multiple jet
production. Produced gluons are ordered in their rapidities
and then connected with their parent valence quarks or diquarks
to form string systems. Finally, fragmentation subroutine of JETSET
is called for hadronization.

\subsection{Cross sections}
\label{sec:jet1}

        In pQCD, the cross section of hard parton scatterings
can be written as\cite{eichten}
\begin{equation}
        \frac{d\sigma_{jet}}{dP_T^2dy_1dy_2} =
        K\sum_{a,b} x_1 x_2 f_a(x_1,P_T^2)f_b(x_2,P_T^2)
        d\sigma^{ab}(\hat{s},\hat{t},\hat{u})/d\hat{t}, \label{eq:sjet1}
\end{equation}
where the summation runs over all parton species, $y_1$,$y_2$
are the rapidities of the scattered partons and $x_1$,$x_2$ are
the fractions of momentum carried by the initial partons and
they are related by $x_1=x_T(e^{y_1}+e^{y_2})/2$,
$x_2=x_T(e^{-y_1}+e^{-y_2})$, $x_T=2P_T/\sqrt{s}$. A factor,
$K\approx 2$ accounts roughly for the higher order corrections.
The default structure functions, $f_a(x,Q^2)$, in HIJING are taken
to be Duke-Owens structure function set 1\cite{duke}.
In future versions some other new parametrizations might be included.

        Integrating Eq.~\ref{eq:sjet1} with a low $P_T$
cutoff $P_0$, we can calculate the total inclusive jet cross
section $\sigma_{jet}$. The average number of semihard parton collisions
for a nucleon-nucleon collision at impact parameter $b$ is
$\sigma_{jet}T_N(b)$, where $T_N(b)$ is partonic overlap function
between the two nucleons. In terms of a semiclassical
probabilistic model\cite{gaisser,wang91a,heureux}, the probability
for multiple minijets production is then
\begin{equation}
        g_j(b)=\frac{[\sigma_{jet}T_N(b)]^j}{j!}e^{-\sigma_{jet}T_N(b)},\;\;
                j\geq 1. \label{eq:sjet3}
\end{equation}
Similarly, we can also represent the soft interactions by
an inclusive cross section $\sigma_{soft}$ which, unlike
$\sigma_{jet}$, can only be determined phenomenologically.
The probability for only soft interactions without any hard
processes is then,
\begin{equation}
        g_0(b)=[1-e^{-\sigma_{soft}T_N(b)}]e^{-\sigma_{jet}T_N(b)}.
                \label{eq:sjet4}
\end{equation}
We have then the total inelastic cross section for nucleon-nucleon
collisions,
\begin{eqnarray}
        \sigma_{in}&=&\int{d^2b}\sum_{j=0}^{\infty}g_j(b) \nonumber \\
        &=&\int{d^2b}[1-e^{-(\sigma_{soft}+\sigma_{jet})T_N(b)}].
                \label{eq:cin}
\end{eqnarray}
Define a real eikonal function,
\begin{equation}
        \chi(b,s)\equiv\frac{1}{2}\sigma_{soft}(s)T_N(b,s)+
                        \frac{1}{2}\sigma_{jet}(s)T_N(b,s), \label{eq:eiko}
\end{equation}
we have the elastic, inelastic, and total cross sections of
nucleon-nucleon collisions,
\begin{equation}
        \sigma_{el}=\pi\int_{0}^{\infty}db^2\left[1-
                e^{-\chi(b,s)}\right]^2, \label{eq:cin1}
\end{equation}
\begin{equation}
        \sigma_{in}=\pi\int_{0}^{\infty}db^2\left[1-
                e^{-2\chi(b,s)}\right],\label{eq:cin2}
\end{equation}
\begin{equation}
        \sigma_{tot}=2\pi\int_{0}^{\infty}db^2\left[1-
                e^{-\chi(b,s)}\right],\label{eq:cin3}
\end{equation}
We assume that the parton density in a nucleon can be
approximated  by the Fourier transform of a dipole form factor.
The overlap function is then,
\begin{equation}
        T_N(b,s)=2\frac{\chi_0(\xi)}{\sigma_{soft}(s)},\label{eq:over1}
\end{equation}
with
\begin{equation}
        \chi_0(\xi)=\frac{\mu_0^2}{96}(\mu_0 \xi)^3 K_3(\mu_0 \xi),
                \;\; \xi=b/b_0(s),\label{eq:over2}
\end{equation}
where $\mu_0=3.9$ and $\pi b_0^2(s)\equiv\sigma_0=\sigma_{soft}(s)/2$
is a measure of the geometrical size of the nucleon. The
eikonal function then can be written as,
\begin{equation}
        \chi(b,s)\equiv\chi(\xi,s)
        =[1+\sigma_{jet}(s)/\sigma_{soft}(s)]\chi_0(\xi).
\end{equation}

        $P_0\simeq 2$ GeV/$c$ and a constant value of
$\sigma_{soft}(s)=57$ mb are chosen to fit the experimental
data on cross sections\cite{wang91a} in $pp$ and $p\bar{p}$
collisions. We shall follow the equations listed above to simulate
multiple jets production at the level of nucleon-nucleon
collisions in HIJING Monte Carlo program. Once the number
of hard scatterings is determined, we then use PYTHIA to generate
the kinetic variables of the scattered partons and the initial
and final state radiations.

\subsection{Jet triggering}
\label{sec:jet2}

Because the differential cross section of jet production
decreases for several orders in magnitude from small to
large $P_T$, we often have to trigger on jet production
with specified $P_T$ in order to increase the simulation
efficiency. The triggering can then change the probability
of multiple minijet production and thus the whole event
structure. In particular, such rare processes of large
$P_T$ scatterings most often occur when the impact
parameter of nucleon-nucleon collision is small so that
the partonic overlap is large. At small impact parameters,
the production of multiple jets is then enhanced.

If we want to trigger on events which have at least one
jet with $P_T$ above $P_T^{trig}$, the conditional
probability for multiple minijet production in the
triggered events is then\cite{hijing},
\begin{equation}
        g_j^{trig}(b) = \frac{[\sigma_{jet}(P_0)T_N(b)]^j}{j!}
        \left\{1-\left[\frac{\sigma_{jet}(P_0)-\sigma_{jet}(P_T^{trig})}
        {\sigma_{jet}(P_0)}\right]^j\right\}e^{-\sigma_{jet}(P_0)T_N(b)}.
                 \label{eq:trigjet4}
\end{equation}
It is obvious that $g_j^{trig}(b)$ returns back to $g_j(b)$
(Eq. \ref{eq:sjet3}) when $P_T^{trig}=P_0$.  Summing over $j\geq 1$
leads to the expected total probability for having at least one
jet with $P_T>P_T^{trig}$,
\begin{equation}
        g^{trig}(b)=1-e^{-\sigma_{jet}(P_T^{trig})T_N(b)},
                 \label{eq:trigjet5}
\end{equation}

        Since $g_j^{trig}(b)$ differs from $g_j(b)$, the
triggering of a particular jet therefore has changed the
production rates of the other jets in the same event.
This triggering effect is especially significant when we
consider large $P_T^{trig}$.  It becomes more probable to
produce multiple jets due to the triggering on a high $P_T$ jet.
In HIJING, we implement Eq.~\ref{eq:trigjet4} by simulating
two Poisson-like multiple jet distributions with inclusive
cross sections  $\sigma_{jet}(P_0)-\sigma_{jet}(P_T^{trig})$
and $\sigma_{jet}(P_T^{trig})$ respectively. We demand that
the second one must have at least one jet and convolute the
two together. The resultant distribution will be the
triggered distribution.

\subsection{Soft interactions}

        Besides the processes with large transverse momentum
transfer which are described by pQCD, there are also many
small $P_T$ exchanges or soft interactions between two colliding
hadrons.  We adopt a variant of the multiple string phenomenological
model for such soft interactions in which multiple soft gluon
exchanges between valence quarks or diquarks lead to longitudinal
string-like excitations. Gluon production from hard processes
and soft radiations are included as kinks in the strings.
The strings then hadronize according to Lund JETSET7.2 fragmentation
scheme.

        In the center of mass frame of two colliding nucleons
with initial light-cone momenta
\begin{equation}
        p_1=(p_1^+,\frac{m_1^2}{p_1^+},{\bf 0}_T),\;\;\;\;
        p_2=(\frac{m_2^2}{p_2^-}, p_2^-,{\bf 0}_T),
\end{equation}
and $(p_1+p_2)^2=s$, the excited strings will have final
momenta
\begin{equation}
        p'_1=(p_1^+ -P^+,\frac{m_1^2}{p_1^+}+P^-, {\bf P}_T),\;\;\;\;
        p'_2=(\frac{m_2^2}{p_2^-}+P^+, p_2^- -P^-,-{\bf P}_T),
\end{equation}
after a collective momentum exchange $P=(P^+,P^-,{\bf P}_T)$.
The soft interactions by definition have small transverse
momentum transfer, $P_T<1$ GeV/$c$, while large effective
light-come momentum\cite{fritiof} exchange can give rise to
two excited strings with large invariant masses. Defining
\begin{equation}
        P^+=x_+\sqrt{s}-\frac{m_2^2}{p_2^-},\;\;\;\;
        P^-=x_-\sqrt{s}-\frac{m_1^2}{p_1^+},
\end{equation}
the excited masses of the two strings will be
\begin{equation}
        M_1^2=x_-(1-x_+)s-P_T^2, \;\;\;\; M_2^2=x_+(1-x_-)s-P_T^2,
                        \label{eq:strnms}
\end{equation}
respectively. If we require that the excited string masses must
have a minimum value $M_{cut}$, then the kinematically
allowed region of $x^{\pm}$
will be
\begin{equation}
        x_{\mp}(1-x_{\pm})\geq M_{Tcut}^2/s, \label{eq:xregn}
\end{equation}
where $M_{Tcut}^2=M_{cut}^2+P_T^2$.
The condition for the above equations to be valid is
\begin{equation}
        \sqrt{s}\geq 2M_{Tcut}. \label{eq:smin}
\end{equation}
This is the minimum colliding energy we will require to produce
two excited strings which can be fragmented into hadrons by the
Lund string fragmentation model. We have chosen $M_{cut}$ to be
1.5 GeV/$c^2$ in all our calculations involving nucleon collisions.
When the energy is smaller than what Eq.~\ref{eq:smin}
requires, we assume that the interaction can be described by
other processes like single diffractive or $N^{\star}$ (or $\rho$,
$K^{\star}$ in cases of pions and kaons collisions) excitation.
However, we usually do not expect that the model is still valid
at such low energies.  Eq.~\ref{eq:smin} also serves as to
determine the maximum $P_T$ that the strings can obtain from
the soft interactions. If hard interactions are  involved,
the kinetic boundary of string formation is reduced by the
hard scatterings.

        In order to best fit the rapidity distributions of charged
particles, we choose the following distributions for light-cone
momentum transfer,
\begin{equation}
        P(x_{\pm})=\frac{(1.0-x_{\pm})^{1.5}}
                {(x_{\pm}^2+c^2/s)^{1/4}},
                \label{eq:xdistr1}
\end{equation}
for nucleons and
\begin{equation}
        P(x_{\pm})=\frac{1}{(x_{\pm}^2+c^2/s)^{1/4}
                [(1-x_{\pm})^2+c^2/s]^{1/4}},
                \label{eq:xdistr2}
\end{equation}
for mesons, where $c=0.1$ GeV is a cutoff for computational
purpose with little theoretical consequences in the model.
For single-diffractive events whose cross section
can be obtained from an empirical parametrization\cite{goulianos},
we fix the mass of the diffractive hadron to be its own or its
vector state excitation and find the mass of the single excited
string according to the well known distribution,
\begin{equation}
        P(x_{\pm})=\frac{1}{(x_{\pm}^2+c^2/s)^{1/2}}, \label{eq:xdistr3}
\end{equation}
which lead to the experimentally observed\cite{goulianos}
mass distribution $dM^2/M^2$ of the disassociated hadrons.

        Before fragmentation, the excited strings are also assumed to
have soft gluon radiation induced by the soft interactions. Such
soft gluon radiation can be approximated by color dipole model
as has been successfully implemented in ARIADNE Monte Carlo
program\cite{dipole}. In HIJING, we adopted subroutines AR3JET
and ARORIE from FRITIOF 1.7\cite{fritiof} to simulate the dipole
radiation which appear as gluon kinks in the string. Since minijets
are treated explicitly via pQCD, we limit the transverse momentum
of the radiated gluons below the minijet cutoff $P_{0}=2$ GeV/$c$.
The limitation on the transverse momentum is a characteristic
feature of induced bremsstrahlung due to soft exchanges\cite{gunion}.
The invariant mass cutoff for strings to radiate is fixed at
$M_{cut}^{rad}=2$ GeV/$c^2$ by default.

\subsection{$P_T$ kick from soft interactions}

        As described in the above, hard or semihard scatterings in our
model have at least transverse momentum of $P_T\geq P_0$. The value of
$P_0$ we use is the result of a model dependent fit of calculated cross
sections to the experimental values. One can imagine that the
corresponding soft interactions, which are characterized by inclusive
cross section $\sigma_{soft}$, will depend on $P_0$. For such
processes, we include an extra low $P_T<P_0$ transfer to the
valence quarks or diquarks at string end points. We assume a
distribution for the $P_T$ kick which extrapolates smoothly to
the high $P_T$ regime of hard scatterings but vary more slowly
for $P_T\ll P_0$,
\begin{equation}
        f_{kick}(P_T)\approx\frac{\theta(P_0-P_T)}{(P_T^2+c^2)
        (P_T^2+P_{0}^2)},\label{eq:kick}
\end{equation}
where $c=0.1$ GeV/$c$. In practice, the distribution will follow
a Gaussian form when $P_T>P_0$. Since diquarks are composites,
we also assume that $P_T$ transfer to a diquark is relatively
suppressed by a form factor with a scale of 1 GeV/$c$.

        This $P_T$ kick to the quarks or diquarks during the soft
interactions will provide an extra increase in transverse momentum
to produced hadrons in order to fit the experimental data at low
energies\cite{hijing}. Otherwise, the transverse momentum from pair
production in the default Lund string fragmentation is not enough
to account for the higher $P_T$ tail in low energy $pp$ collisions.

\section{Parton Production in $pA$ and $AA$ Collisions}

        To include the nuclear effects on jet production and
fragmentation, we also consider the EMC\cite{shadow1}
effect of the parton structure functions in nuclei and the
interaction of the produced jets with the excited nuclear
matter in heavy ion collisions.

\subsection{Binary approximation and initial state interaction}

        We assume that a nucleus-nucleus collision can be decomposed
into binary nucleon-nucleon collisions which generally involve the
wounded nucleons. In a string picture, the wounded nucleons become strings
excited along the beam direction. At high energy, the excited strings
are assumed to interact again like the ordinary nucleon-nucleon
collisions before they fragment. Unlike FRITIOF model, we allow
an excited string to be de-excited within the kinematic limits
in the subsequent collisions.
The binary approximation can also be applied to rare hard
scatterings which involve only independent pairs of partons. The
probability for a given parton to suffer multiple high $P_T$
scatterings is small and is not implemented in the current
version of the program. We employ a three-parameter Wood-Saxon
nuclear density to compute the number of binary collisions
at a given impact parameter.

        For each one of these binary collisions, we use
the eikonal formalism as given in Section~\ref{sec:jet1}
to  determine the probability of collision, elastic or
inelastic and the number of jets it produces.  After simulation
of hard processes, the energy of the scattered partons is
subtracted from the nucleon and the remaining energy is used
in the soft interaction as in ordinary soft nucleon-nucleon
collisions. The excited string system minus the scattered
partons suffers further collisions according to the geometrical
probability.

        We assign one of the two scattered partons per hard
scattering to each participating nucleon or they may form an
independent single ($q-\bar{q}$) string system.  After all
binary collisions are processed, we then connect the scattered
partons in the associated nucleons with the corresponding
valence quarks and diquarks to form string systems. The strings
are then fragmented into particles.

\subsection{Nuclear shadowing effect}

        One of the most important nuclear effects in relativistic
heavy ion collisions is the nuclear modification of parton
structure functions. It has been observed\cite{shadow1}
that the effective number of quarks and antiquarks in a
nucleus is depleted in the low region of $x$. Though gluon
shadowing has not been studied experimentally, we  will assume
that the shadowing effect for gluons and quarks is the same.
We also neglect the QCD evolution of the shadowing effect in the
current version.  There is no experimental evidence for significant
$Q$ dependence of the nuclear effect on the quark structure functions.
However, theoretical study\cite{eskola93} shows that gluon shadowing may
evolve with $Q$.

        At this stage, the experimental data unfortunately
can not fully determine the $A$ dependence of the shadowing.
We will follow the $A$ dependence as proposed in
Ref.\cite{shadow2} and use the following parametrization,
\begin{eqnarray}
        R_A(x)&\equiv&\frac{f_{a/A}(x)}{Af_{a/N}(x)} \nonumber\\
&=&1+1.19\ln^{1/6}\!A\,[x^3-1.5(x_0+x_L)x^2+3x_0x_Lx]\nonumber\\
                & &-[\alpha_A-\frac{1.08(A^{1/3}-1)}{\ln(A+1)}\sqrt{x}]
                        e^{-x^2/x_0^2},\label{eq:shadow}\\
        \alpha_A&=&0.1(A^{1/3}-1),\label{eq:shadow1}
\end{eqnarray}
where $x_0=0.1$ and $x_L=0.7$. The term proportional to $\alpha_A$ in
Eq.~\ref{eq:shadow} determines the shadowing for $x<x_0$ with the
most important nuclear dependence, while the rest gives the overall
nuclear effect on the structure function in $x>x_0$ with some very slow
$A$ dependence. This parametrization can fit the overall nuclear
effect on the quark structure function in the small and medium
$x$ region\cite{hijing}.

        To take into account of the impact parameter dependence,
we assume that the shadowing effect $\alpha_A$ is proportional
to the longitudinal dimension of the nucleus along the straight
trajectory of the interacting nucleons. We thus parametrize
$\alpha_A$ in Eq.~\ref{eq:shadow} as
\begin{equation}
        \alpha_A(r)=0.1(A^{1/3}-1)\frac{4}{3}\sqrt{1-r^2/R_A^2},
                        \label{eq:rshadow}
\end{equation}
where $r$ is the transverse distance of the interacting
nucleon from its nucleus center and $R_A$ is the radius of the
nucleus. For a sharp sphere nucleus with overlap function
$T_A(r)=(3A/2\pi R_A^2)\sqrt{1-r^2/R_A^2}$, the averaged
$\alpha_A(r)$ is $\alpha_A=\pi\int_0^{R_A^2}dr^2 T_A(r)\alpha_A(r)/A$.
Because the rest of Eq.~\ref{eq:shadow} has a very slow $A$
dependence, we will only consider the impact parameter dependence
of $\alpha_A$. After all, most of the jet productions occur in
the small $x$ region where only shadowing is important.

        To simplify the calculation during the Monte Carlo
simulation, we can decompose $R_A(x,r)$ into two parts,
\begin{equation}
        R_A(x,r)\equiv R_A^0(x)-\alpha_A(r)R_A^s(x),
\end{equation}
where $\alpha_A(r)R_A^s(x)$ is the term proportional to $\alpha_A(r)$
in Eq.~\ref{eq:shadow} with  $\alpha_A(r)$ given in Eq.~\ref{eq:rshadow}
and $R_A^0(x)$ is the rest of $R_A(x,r)$. Both $R_A^0(x)$ and $R_A^s(x)$
are now independent of $r$. The effective jet production cross section
of a binary nucleon-nucleon  interaction in $A+B$ nuclear collisions
is then,
\begin{equation}
        \sigma_{jet}^{eff}(r_A,r_B)=\sigma_{jet}^0-\alpha_A(r_A)\sigma_{jet}^A
                -\alpha_B(r_B)\sigma_{jet}^B
                +\alpha_A(r_A)\alpha_B(r_B)\sigma_{jet}^{AB},\label{eq:sjetab}
\end{equation}
where $\sigma_{jet}^0$, $\sigma_{jet}^A$, $\sigma_{jet}^B$ and
$\sigma_{jet}^{AB}$ can be calculated through Eq.~\ref{eq:sjet1} by
multiplying \\
$f_a(x_1,P_T^2)f_b(x_2,P_T^2)$ in the integrand with
$R_A^0(x_1)R_B^0(x_2)$, $R_A^s(x_1)R_B^0(x_2)$, $R_A^0(x_1)R_B^s(x_2)$ and
$R_A^s(x_1)R_B^s(x_2)$ respectively. With calculated values of
$\sigma_{jet}^0$, $\sigma_{jet}^A$, $\sigma_{jet}^B$ and $\sigma_{jet}^{AB}$,
we will know the effective jet cross section
$\sigma_{jet}^{eff}$ for any binary nucleon-nucleon collision.

\subsection{Final state parton interaction}

        Another important nuclear effect on the jet
production in heavy ion collisions is the final state
integration. In  high energy heavy ion collisions, a dense
hadronic or partonic matter must be produced in the central
region. Because this matter can extend over a transverse
dimension of at least $R_A$, jets with large $P_T$ from
hard scatterings have to traverse this hot environment. For
the purpose of studying the property of the dense matter
created during the nucleus-nucleus collisions, it is
important to investigate the interaction of jets with the
matter and the energy loss they suffer during their
journey out. It is estimated\cite{gyu92,dedx} that the gluon
bremsstrahlung induced by soft interaction dominate the
energy loss mechanism.

        We model the induced radiation in HIJING via a simple
collinear gluon splitting scheme with given energy loss $dE/dz$.
The energy loss for gluon jets is twice that of quark jets\cite{dedx}.
We assume that interaction only occur with the locally comoving
matter in the transverse direction. The interaction points are
determined via a probability
\begin{equation}
        dP=\frac{d\ell}{\lambda_s}e^{-\ell/\lambda_s},
\end{equation}
with given mean free path $\lambda_s$, where $\ell$ is the
distance the jet has traveled after its last interaction.
The induced radiation is simulated by transferring a part of
the jet energy $\Delta E(\ell)=\ell dE/z$ as a gluon kink to
the other string which the jet interacts with. We continue
the procedure until the jet is out of the whole excited system
or when the jet energy is smaller than a cutoff below which
a jet can not loss energy any more. We take this cutoff as
the same as the cutoff $P_0$ for jet production. To determine
how many and which excited strings could interact with the
jet, we also have to assume a cross section of jet interaction so that
excited strings within a cylinder of radius $r_s$ along the jet
direction could interact with the jet. $\lambda_s$
should be related to $r_s$ via the density of the system of excited
strings. We simply take them as two parameters in our model.

\section{Program Description}

        HIJING 1.0, written in  FORTRAN 77 is a Monte Carlo
simulation package for parton and particle production
in high energy hadron-hadron, hadron-nucleus,
and nucleus-nucleus collisions. It consists of subroutines for
physics simulation and common blocks for parameters and event
records. Users have to provide their own main program where desired
parameters and event type are specified, and simulated events
can be studied. HIJING 1.0 uses PYTHIA 5.3 to generate kinetic
variables for each hard scattering and JETSET 7.2 for jet
fragmentation. Therefore, HIJING 1.0 uses the same particle
flavor code (included in the appendix)
as JETSET 7.2 and PYTHIA 5.3. Users can also
obtain more flexibility by using subroutines in JETSET 7.2 and
changing the values of parameters in JETSET 7.2 and PYTHIA 5.3
therein. We refer users to the original literature\cite{pythia,jetset}
for the documentations of JETSET 7.2 and PYTHIA 5.3.
For many users, however, subroutines, parameters and event
information in HIJING 1.0 alone will be enough for studying
most of the event types and the physics therein. To save compiling
time and to meet some specific needs of HIJING 1.0, PYTHIA 5.3 has
been modified and together with JETSET 7.2 is renamed as HIPYSET.
Therefore, one should link the main program with HIJING and HIPYSET.

        In this program, implicit integer numbers are assumed for
variables beginning with letters I--M, while the implicit real numbers
are assumed for variables beginning with letters A--H and O--Z.

\subsection{Random numbers}

        Random numbers in HIJING is obtained by calling the
VAX VMS system function RAN(NSEED). On SPARCstation, one has to
link the program with {\tt -xl} flag in order to compile the program.
We have not checked the portability of this function on machines
with other operating systems. Whenever one encounters problem with
this (pseudo) random number generator on different machines other
than VAX and SPARCstation, one should replace this function by
another random number generator.

        To start a new sequence of random numbers, one should give
a new large odd integer value to variable NSEED in
COMMON/RANSEED/NSEED.

\subsection{Main subroutines}

        After supplying the desired parameters, the first subroutine a
user has to call is HIJSET. Then subroutine HIJING can be called to
simulate the specified events.

\begin{description}
\itemsep=-4.0pt
\item{}SUBROUTINE  HIJSET (EFRM, FRAME, PROJ, TARG, IAP, IZP, IAT, IZT)
\item{}Purpose: to initialize HIJING for specified event type, collision
                frame and energy.
\item{}EFRM: colliding energy (GeV) per nucleon in the frame specified
                by FRAME.
\item{}FRAME: character variable to specify the frame of the collision.
        \vspace{-12.0pt}
        \begin{description}
        \itemsep=-4.0pt
                \item{}='CMS': nucleon-nucleon center of mass frame,
                        with projectile momentum in $+z$ direction and
                        target momentum in $-z$ direction.
                \item{}='LAB': laboratory frame of the fixed target with
                        projectile momentum in $+z$ direction.
        \end{description}
        \vspace{-4.0pt}
\item{}PROJ, TARG: character variables of projectile and target particles.
        \vspace{-12.0pt}
        \begin{description}
        \itemsep=-4.0pt
                \item{}='P': proton.
                \item{}='PBAR': anti-proton.
                \item{}='N': neutron.
                \item{}='NBAR': anti-neutron.
                \item{}='PI$+$': $\pi^+$.
                \item{}='PI$-$': $\pi^-$.
                \item{}='A': nucleus.
        \end{description}
        \hspace{-4.0pt}
\item{}IAP, IAT: mass number of the projectile and target nucleus. Set
                to 1 for hadrons.
\item{}IZP, IZT: charge number of the projectile and target nucleus, for
                hadrons it is the charge number of that hadron (=1, 0, $-1$).
\end{description}

\begin{description}
\itemsep=-4.0pt
\item{}SUBROUTINE  HIJING (FRAME, BMIN, BMAX)
\item{}Purpose: to generate a complete event as specified by subroutine HIJSET
                and the given parameters as will be described below.
                This is the main routine which can be called (many times)
                only after HIJSET has been called once.
\item{}FRAME: character variable to specify the frame of the collision
                as given in the HIJSET call.
\item{}BMIN, BMAX: low and up limits (fm) between which the impact
                parameter squared $b^2$ is uniformly distributed
                for $pA$ and
                $AB$ collisions. For hadron-hadron collisions, both are set
                to zero and the events
                are automatically averaged over all impact parameters.
\end{description}

\subsection{Common blocks for event information}

        There are mainly three common blocks which provide users with
important information of the generated events. Common block HIMAIN1
contains global information of the events and common block HIMAIN2
of the produced particles. The information of produced partons are
stored in common blocks HIJJET1, HIJJET2, HISTRNG.

\begin{description}
\itemsep=-4.0pt
\item{}COMMON/HIMAIN1/NATT, EATT, JATT, NT, NP, N0, N01, N10, N11
\item{}Purpose: to give the overall information of the generated event.
\item{}NATT: total number of produced stable and undecayed particles of
                the current event.
\item{}EATT: the total energy of the produced particles in c.m. frame
                of the collision to check energy conservation.
\item{}JATT: the total number of hard scatterings in the current event.
\item{}NP, NT: the number of participant projectile and target nucleons
                in the current event.
\item{}N0, N01, N10, N11: number of $N$-$N$, $N$-$N_{wounded}$,
                $N_{wounded}$-$N$, and
                $N_{wounded}$-$N_{wounded}$ collisions in
                the current event ($N$, $N_{wounded}$ stand
                for nucleon and wounded nucleon respectively).
\end{description}

\begin{description}
\itemsep=-4.0pt
\item{}COMMON /HIMAIN2/KATT(130000,4), PATT(130000,4)
\item{}Purpose: to give information of produced stable and undecayed
                particles. Parent particles which decayed are not included
                here.
\item{}KATT(I, 1): (I=1,$\cdots$,NATT) flavor codes (see appendix) of
                the produced particles.
\item{}KATT(I, 2): (I=1,$\cdots$,NATT) status codes to identify the
                sources from which the particles come.
        \vspace{-12.0pt}
        \begin{description}
        \itemsep=-4.0pt
                \item{}=0: projectile nucleon (or hadron) which has
                        not interacted at all.
                \item{}=1: projectile nucleon (or hadron) which
                        only suffers an elastic collision.
                \item{}=2: from a diffractive projectile nucleon (or hadron)
                        in a single diffractive interaction.
                \item{}=3: from the fragmentation of a projectile string
                        system (including gluon jets).
                \item{}=10 target nucleon (or hadron) which has not
                        interacted at all.
                \item{}=11: target nucleon (or hadron) which only
                        suffers an elastic collision.
                \item{}=12: from a diffractive target nucleon (or hadron)
                        in a single diffractive interaction.
                \item{}=13: from the fragmentation of a target string
                        system (including gluon jets).
                \item{}=20: from scattered partons which form string
                        systems themselves.
                \item{}=40: from direct production in the hard processes
                        ( currently, only direct photons are included).
        \end{description}
        \vspace{-4.0pt}
\item{}KATT(I,3): (I=1,$\cdots$,NATT) line number of the parent particle.
                        For finally produced or directly produced (not from
                        the decay of another particle) particles, it is set
                        to 0 (The option to keep the information of all
                        particles including the decayed ones is IHPR2(21)=1).
\item{}KATT(I,4): (I=1,$\cdots$,NATT) status number of the particle.
        \vspace{-12.0pt}
        \begin{description}
        \itemsep=-4.0pt
               \item{}=1: finally or directly produced particles.
               \item{}=11: particles which has already decayed.
        \end{description}
        \vspace{-4.0pt}
\item{}PATT(I, 1-4): (I=1,$\cdots$,NATT) four-momentum ($p_x,p_y,p_z,E$)
                (GeV/$c$, GeV) of the produced particles.
\end{description}

\begin{description}
\itemsep=-4.0pt
\item{}COMMON/HIJJET1/NPJ(300), KFPJ(300,500), PJPX(300,500), PJPY(300,500),\\
PJPZ(300,500), PJPE(300,500), PJPM(300,500), NTJ(300), KFTJ(300,500),\\
PJTX(300,500), PJTY(300,500), PJTZ(300,500), PJTE(300,500), PJTM(300,500)
\item{}Purpose: contains information about produced partons which are
                connected with the valence quarks and diquarks of
                projectile or target nucleons (or hadron) to form
                string systems for fragmentation. The momentum and
                energy of all produced partons are calculated in
                the c.m. frame of the collision. IAP, IAT are the
                numbers of nucleons in projectile and target nucleus
                respectively (IAP, IAT=1 for hadron projectile or target).
\item{}NPJ(I): (I=1,$\cdots$,IAP) number of partons associated with projectile
                nucleon I.
\item{}KFPJ(I, J): (I=1,$\cdots$,IAP, J=1,$\cdots$,NPJ(I)) parton
                flavor code of the
                parton J associated with projectile nucleon I.
\item{}PJPX(I, J), PJPY(I, J), PJPZ(I, J), PJPE(I, J), PJPM(I, J): the four
                momentum and mass ($p_x,p_y,p_z,E,M$)
                (GeV/$c$, GeV, GeV/$c^2$) of parton J associated with
                the projectile nucleon I.
\item{}NTJ(I): (I=1,$\cdots$,IAT) number of partons associated with
                target nucleon I.
\item{}KFTJ(I, J): (I=1,$\cdots$,IAT, J=1,$\cdots$,NTJ(I)): parton
                flavor code of the  parton J associated with
                target nucleon I.
\item{}PJTX(I, J), PJTY(I, J), PJTZ(I, J), PJTE(I, J), PJTM(I, J): the four
                momentum and mass ($p_x,p_y,p_z,E,M$)
                (GeV/$c$, GeV, GeV/$c^2$) of parton J associated with
                target nucleon I.
\end{description}

\begin{description}
\itemsep=-4.0pt
\item{}COMMON/HIJJET2/NSG, NJSG(900), IASG(900,3), K1SG(900,100),\\
\hspace{-24pt}K2SG(900,100), PXSG(900,100), PYSG(900,100), PZSG(900,100), \\
\hspace{-24pt}PESG(900,100), PMSG(900,100)
\item{}Purpose: contains information about the produced partons which
                will form string systems themselves without being
                connected to valence quarks and diquarks.
\item{}NSG: the total number of such string systems.
\item{}NJSG(I): (I=1,$\cdots$,NSG) number of partons in the string system I.
\item{}IASG(I, 1), IASG(I, 2): to specify which projectile and target
                nucleons produce string system I.
\item{}IASG(I, 3): to indicate whether the jets will be quenched (0)
                or will not be quenched (1).
\item{}K1SG(I, J): (J=1,$\cdots$,NJSG(I)) color flow information of parton J
                in string system I (see JETSET 7.2 for detailed
                explanation).
\item{}K2SG(I, J): (J=1,$\cdots$,NJSG(I)) flavor code of parton J in string
                system I.
\item{}PXSG(I, J), PYSG(I, J), PZSG(I, J), PESG(I, J), PMSG(I, J): four
                momentum and mass ($p_x,p_y,p_z,E,M$)
                ( GeV/$c$, GeV, GeV/$c^2$) of parton J in string system I.
\end{description}

\begin{description}
\itemsep=-4.0pt
\item{}COMMON/HISTRNG/NFP(300,15), PP(300,15), NFT(300,15), PT(300,15)
\item{}Purpose: contains information about the projectile and
        target nucleons (hadron) and the corresponding constituent
        quarks, diquarks. IAP, IAT are the numbers of nucleons in
        projectile and target nucleus respectively (IAP, IAT=1
        for hadron projectile or target).
\item{}NFP(I, 1): (I=1,$\cdots$,IAP) flavor code of the valence quark in
                projectile nucleon (hadron) I.
\item{}NFP(I, 2): flavor code of diquark in projectile nucleon (anti-quark
                in projectile meson) I.
\item{}NFP(I, 3): present flavor code of the projectile nucleon (hadron) I
                ( a nucleon or meson can be excited to its vector resonance).
\item{}NFP(I, 4): original flavor code of projectile nucleon (hadron) I.
\item{}NFP(I, 5): collision status of projectile nucleon (hadron) I.
        \vspace{-12.0pt}
        \begin{description}
        \itemsep=-4.0pt
                \item{}=0: suffered no collision.
                \item{}=1: suffered an elastic collision.
                \item{}=2: being the diffractive one in a single-diffractive
                        collision.
                \item{}=3: became an excited string after an inelastic
                        collision.
        \end{description}
        \vspace{-4.0pt}
\item{}NFP(I, 6): the total number of hard scatterings associated with
                projectile nucleon (hadron) I. If NFP(I,6)$<0$, it can not
                        produce jets any more due to energy conservation.
\item{}NFP(I, 10): to indicate whether the valence quarks or diquarks
                (anti-quarks) in projectile nucleon (hadron) I
                suffered a hard scattering,
        \vspace{-12.0pt}
        \begin{description}
        \itemsep=-4.0pt
                \item{}=0: has not  suffered a hard scattering.
                \item{}=1: suffered one or more hard scatterings in
                        current binary nucleon-nucleon collision.
                \item{}=$-1$: suffered one or more hard scatterings in
                        previous binary nucleon-nucleon collisions.
        \end{description}
        \vspace{-4.0pt}
\item{}NFP(I, 11): total number of interactions projectile nucleon (hadron)
                I  has suffered so far.
\item{}PP(I, 1), PP(I, 2), PP(I, 3), PP(I, 4), PP(I, 5): four momentum and
                the invariant mass ($p_x,p_y,p_z,E,M$)
                (GeV/$c$, GeV, GeV/$c^2$) of projectile nucleon (hadron) I.
\item{}PP(I, 6), PP(I, 7): transverse momentum ($p_x,p_y$) (GeV/$c$) of the
                valence quark in projectile nucleon (hadron) I.
\item{}PP(I, 8), PP(I, 9): transverse momentum ($p_x,p_y$) (GeV/$c$) of the
                diquark (anti-quark) in projectile nucleon (hadron) I.
\item{}PP(I, 10), PP(I, 11), PP(I, 12): three momentum ($p_x,p_y,p_z$)
                (GeV/$c$) transferred to the quark or diquark (anti-quark)
                in projectile nucleon (hadron) I from the last hard
                scattering.
\item{}PP(I, 14): mass (GeV/$c^2$) of the quark in projectile nucleon
                (hadron) I.
\item{}PP(I, 15): mass of the diquark (anti-quark) in projectile
                nucleon (hadron) I.
\item{}NFT(I, 1--15), PT(I,1--15): give the same
                information for the target nucleons (hadron) and the
                corresponding quarks and diquarks (anti-quarks) as for
                the projectile nucleons.
\end{description}

\subsection{Options and parameters}

        The following common block is for input parameters for HIJING
which are used mainly for specifying event options and changing the
default parameters. It also contains some extra event information.
The default values of the parameters are given by D. Some parameters
are simply used to redefine the parameters in JETSET 7.2 and PYTHIA 5.3.
Users have to find the detailed explanations in JETSET and PYTHIA
documentations.

\begin{description}
\itemsep=-4.0pt
\item{}COMMON/HIPARNT/HIPR1(100), IHPR2(50), HINT1(100), IHNT2(50)
\item{}Purpose: contains input parameters (HIPR1, IHPR2) for event options
                and some extra information (HINT1, IHNT2) of current event.
\item{}HIPR1(1): (D=1.5 GeV/$c^2$) minimum value for the invariant mass of
                the excited string system in a hadron-hadron interaction.
\item{}HIPR1(2): (D=0.35 GeV) width of the Gaussian $P_T$ distribution of
                produced hadron in Lund string fragmentation
                (PARJ(21) in JETSET 7.2).
\item{}HIPR1(3), HIPR1(4): (D=0.5, 0.9 GeV$^{-2}$) give the $a$ and $b$
                parameters of the symmetric Lund fragmentation function
                (PARJ(41), PARJ(42) in JETSET 7.2).
\item{}HIPR1(5): (D=2.0 GeV/$c^2$) invariant mass cut-off for the dipole
                radiation of a string system below which soft gluon
                radiations are terminated.
\item{}HIPR1(6): (D=0.1) the depth of shadowing of structure functions
                at $x=0$ as defined in Eq.~\ref{eq:shadow1}:
                $\alpha_A=\mbox{HIPR1(6)}\times(A^{1/3}-1)$.
\item{}HIPR1(7): not used
\item{}HIPR1(8): (D=2.0 GeV/$c$) minimum $P_T$ transfer in hard or
                semihard scatterings.
\item{}HIPR1(9): (D=$-1.0$ GeV/$c$) maximum $P_T$ transfer in hard or
                semihard scatterings. If negative, the limit is set
                by the colliding energy.
\item{}HIPR1(10): (D=$-2.25$ GeV/$c$) specifies the value of $P_T$ for
                each triggered hard scattering generated per event
                (see Section \ref{sec:jet2}). If HIPR1(10) is negative,
                its absolute value gives the low limit of the
                $P_T$ of the triggered jets.
\item{}HIPR1(11): (D=2.0 GeV/$c$) minimum $P_T$ of a jet which will interact
                with excited nuclear matter. When the $P_T$ of a jet
                is smaller than HIPR1(11) it will stop interacting further.
\item{}HIPR1(12): (D=1.0 fm) transverse distance between a traversing jet
                and an excited nucleon (string system) below which they
                will interact and the jet will lose energy and momentum
                to that string system.
\item{}HIPR1(13): (D=1.0 fm) the mean free path of a jet when it goes
                through the excited nuclear matter.
\item{}HIPR1(14): (D=2.0 GeV/fm) the energy loss $dE/dz$ of a gluon
                jet inside the excited nuclear matter. The energy loss
                for a quark jet is half of the energy loss of a gluon.
\item{}HIPR1(15): (D=0.2 GeV/$c$) the scale $\Lambda$ in the
                calculation of $\alpha_s$.
\item{}HIPR1(16): (D=2.0 GeV/$c$) the initial scale $Q_0$ for the
                evolution of the structure functions.
\item{}HIPR1(17): (D=2.0) $K$ factor for the differential jet cross
                sections in the lowest order pQCD calculation.
\item{}HIPR1(18): not used
\item{}HIPR1(19), HIPR1(20): (D=0.1, 1.4 GeV/$c$) parameters in the
                distribution for the $P_T$ kick from soft interactions
                (see Eq.~\ref{eq:kick}),
                $1/[(\mbox{HIPR1(19)}^2+P_T^2)(\mbox{HIPR1(20)}^2+P_T^2)]$.
\item{}HIPR1(21): (D=1.6 GeV/$c$) the maximum $P_T$ for soft interactions,
                beyond which a Gaussian distribution as specified by
                HIPR1(2) will be used.
\item{}HIPR1(22): (D=2.0 GeV/$c$) the scale in the form factor to suppress
                the $P_T$ transfer to diquarks in hard scatterings,
\item{}HIPR1(23)--HIPR1(28): not used.
\item{}HIPR1(29): (D=0.4 fm) the minimum distance between two nucleons
                inside a nucleus when the coordinates of the nucleons
                inside a nucleus are initialized.
\item{}HIPR1(30): (D=2$\times$HIPR1(31)=57.0 mb) the inclusive cross
                section $\sigma_{soft}$ for soft interactions. The default
                value $\sigma_{soft}=2\sigma_0$ is used to ensure the
                geometrical scaling of $pp$ interaction cross sections
                at low energies.
\item{}HIPR1(31): (D=28.5 mb) the cross section $\sigma_0$ which
                characterizes the geometrical size of a nucleon
                ($\pi b_0^2=\sigma_0$, see Eq.~\ref{eq:over2}).
                The default value is only for high-energy
                limit ($\sqrt{s}>200$ GeV). At lower energies, a slight
                decrease which depends on energy is parametrized in the
                program. The default values of the two parameters
                HIPR1(30), HIPR1(31) are only for $NN$ type interactions.
                For other kinds of projectile or target hadrons, users
                should change these values so that correct inelastic
                and total cross sections (HINT1(12), HINT1(13)) are
                obtained by the program.
\item{}HIPR1(32): (D=3.90) parameter $\mu_0$ in Eq.~\ref{eq:over2} for
                the scaled eikonal function.
\item{}HIPR1(33): fractional cross section of single-diffractive
                interaction as parametrized in Ref.~\cite{goulianos}.
\item{}HIPR1(34): maximum radial coordinate for projectile nucleons
                to be given by the initialization program HIJSET.
\item{}HIPR1(35): maximum radial coordinate for target nucleons
                to be given by the initialization program HIJSET.
\item{}HIPR1(36)-HIPR1(39): not used.
\item{}HIPR1(40): (D=3.141592654) value of $\pi$.
\item{}HIPR1(41)--HIPR1(42): not used.
\item{}HIPR1(43): (D=0.01) fractional energy error relative to the
                colliding energy permitted per nucleon-nucleon collision.
\item{}HIPR1(44), HIPR1(45), HIPR1(46): (D=1.5, 0.1 GeV, 0.25) parameters
                $\alpha$, $c$ and $\beta$ in the valence quark
                distributions for soft string excitation,
                $(1-x)^{\alpha}/(x^2+c^2/s)^{\beta}$ for baryons,
                $1/{(x^2+c^2/s)[(1-x)^2+c^2/s)]}^{\beta}$ for mesons.
\item{}HIPR1(47), HIPR1(48): (D=0.0, 0.5) parameters $\alpha$ and $\beta$
                in valence quark distribution,
                $(1-x)^{\alpha}/(x^2+c^2/s)^{\beta}$, for the
                disassociated excitation in a single diffractive collision.
\item{}HIPR1(49)--HIPR1(100): not used.
\item{}IHPR2(1): (D=1) switch for dipole-approximated QCD radiation
                of the string system in soft interactions.
\item{}IHPR2(2): (D=3) option for initial and final state radiation in
                the hard scattering.
        \vspace{-12.0pt}
        \begin{description}
        \itemsep=-4.0pt
                \item{}=0: both initial and final radiation are off.
                \item{}=1: initial radiation on and final radiation off.
                \item{}=2: initial radiation off and final radiation on.
                \item{}=3: both initial and final radiation are on.
        \end{description}
        \vspace{-4.0pt}
\item{}IHPR2(3): (D=0) switch for triggered hard scattering with specified
                $P_T\geq$HIPR1(10).
        \vspace{-12.0pt}
        \begin{description}
        \itemsep=-4.0pt
                \item{}=0: no triggered jet production.
                \item{}=1: ordinary hard processes.
                \item{}=2: only direct photon production.
        \end{description}
        \vspace{-4.0pt}
\item{}IHPR2(4): (D=1) switch for jet quenching in the excited
                nuclear matter.
\item{}IHPR2(5): (D=1) switch for the $P_T$ kick due to soft interactions.
\item{}IHPR2(6): (D=1) switch for the nuclear effect on the parton
                distribution function such as shadowing.
\item{}IHPR2(7): (D=1) selection of Duke-Owens set (1 or 2) of parametrization
                of nucleon structure functions.
\item{}IHPR2(8): (D=10) maximum number of hard scatterings per
                nucleon-nucleon interaction. When IHPR2(8)=0, jet
                production will be turned off. When IHPR2(8)$<0$, the
                number of jet production will be fixed at its absolute
                value for each NN collision.
\item{}IHPR2(9): (D=0) switch to guarantee at least one pair of minijets
                production per event ($pp$, $pA$ or $AB$).
\item{}IHPR2(10): (D=0) option to print warning messages about errors that
                might happen. When a fatal error happens the current event
                will be abandoned and a new one is generated.
\item{}IHPR2(11): (D=1) choice of baryon production model.
        \vspace{-12.0pt}
        \begin{description}
        \itemsep=-4.0pt
                \item{}=0: no baryon-antibaryon pair production, initial
                        diquark treated as a unit.
                \item{}=1: diquark-antidiquark pair production allowed,
                        initial diquark treated as a unit.
                \item{}=2: diquark-antidiquark pair production allowed,
                        with the possibility for diquark to split
                        according to the ``popcorn'' scheme (see the
                        documentation of JETSET 7.2).
        \end{description}
        \vspace{-4.0pt}
\item{}IHPR2(12): (D=1) option to turn off the automatic decay of the
                 following particles:
                $\pi^0$, $K^0_S$, $D^{\pm}$, $\Lambda$, $\Sigma^{\pm}$.
\item{}IHPR2(13): (D=1) option to turn on single diffractive reactions.
\item{}IHPR2(14): (D=1) option to turn on elastic scattering.
\item{}IHPR2(15) -- IHPR2(17): not used.
\item{}IHPR2(18):(D=0)option to switch on B-quark production. Charm
                  production is the default. When B-quark production is
                  on, charm quark production is automatically off.
\item{}IHPR2(19): (D=1) option to turn on initial state soft interaction.
\item{}IHPR2(20): (D=1) switch for the final fragmentation.
\item{}IHPR2(21): (D=0) option to keep the information of all particles
                  including those which have decayed and the decay history
                  in the common block HIMAIN2. The line number of the parent
                  particle is KATT(I,3). The status of a partcile,
                  whether it is a finally produced particle (KATT(I,4)=1)
                  or a decayed particle (KATT(I,4)=11) is also kept.
\item{}IHPR2(22)-IHPR2(50): not used.
\item{}HINT1(1): (GeV) colliding energy in the c.m. frame of nucleon-nucleon
                collisions.
\item{}HINT1(2): Lorentz transformation variable $\beta$ from laboratory
                to c.m.  frame of nucleon nucleon collisions.
\item{}HINT1(3): rapidity $y_{cm}$ of the c.m. frame
                $\beta=\tanh y_{cm}$.
\item{}HINT1(4): rapidity of projectile nucleons (hadron) $y_{proj}$.
\item{}HINT1(5): rapidity of target nucleons (hadron) $y_{targ}$.
\item{}HINT1(6): (GeV) energy of the projectile nucleons (hadron) in the
                given frame.
\item{}HINT1(7): (GeV) energy of the target nucleons (hadron) in the
                given frame.
\item{}HINT1(8): (GeV) the rest mass of projectile particles.
\item{}HINT1(9): (GeV) the rest mass of target particles.
\item{}HINT1(10): (mb) the averaged cross section for jet production
                per nucleon-nucleon collisions,
                $\int d^2b\{1-\exp[-\sigma_{jet}T_N(b)]\}$.
\item{}HINT1(11): (mb) the averaged inclusive cross section $\sigma_{jet}$
                for jet production per nucleon-nucleon collisions.
\item{}HINT1(12): (mb) the averaged inelastic cross section of
                nucleon-nucleon collisions.
\item{}HINT1(13): (mb) the averaged total cross section of nucleon-nucleon
                collisions.
\item{}HINT1(14): (mb) the jet production cross section without nuclear
                shadowing effect $\sigma_{jet}^0$ (see Eq.~\ref{eq:sjetab}).
\item{}HINT1(15): (mb) the cross section $\sigma_{jet}^A$ to account for
                the projectile shadowing correction term in the jet cross
                section (see Eq.~\ref{eq:sjetab}).
\item{}HINT1(16): (mb) the cross section $\sigma_{jet}^B$ to account for
                the target shadowing correction term in the jet cross
                section (see Eq.~\ref{eq:sjetab}).
\item{}HINT1(17): (mb) the cross section $\sigma_{jet}^{AB}$ to account
                for the cross term of shadowing correction in the jet
                cross section (see Eq.~\ref{eq:sjetab}).
\item{}HINT1(18): (mb) the effective cross section
                $\sigma_{jet}^{eff}(r_A,r_B)$ for jet production
                of the latest nucleon-nucleon collision which depends
                on the transverse coordinates of the colliding
                nucleons (see Eq.~\ref{eq:sjetab}).
\item{}HINT1(19): (fm) the (absolute value of) impact parameter of the
                latest event.
\item{}HINT1(20): (radians) the azimuthal angle $\phi$ of the impact
                parameter vector in the transverse plane of the latest
                event.
\item{}HINT1(21)--HINT1(25): the four momentum and mass ($p_x,p_y,p_z,E,M$)
                (GeV/$c$, GeV, GeV/$c^2$) of the first scattered parton
                in the triggered hard scattering. This is before the final
                state radiation but after the initial state radiation.
\item{}HINT1(26)--HINT1(30): not used.
\item{}HINT1(31)--HINT1(35): the four momentum and mass ($p_x,p_y,p_z,E,M$)
                (GeV/$c$, GeV, GeV/$c^2$) of the second scattered parton
                in the triggered hard scattering. This is before the final
                state radiation but after the initial state radiation.
\item{}HINT1(46)--HINT1(40): not used.
\item{}HINT1(41)--HINT1(45): the four momentum and mass ($p_x,p_y,p_z,E,M$)
                (GeV/$c$, GeV, GeV/$c^2$) of the first scattered parton
                in the latest hard scattering of the latest event.
\item{}HINT1(46): $P_T$ (GeV/$c$) of the first scattered parton in the
                latest hard scattering of the latest event.
\item{}HINT1(47)--HINT1(50): not used.
\item{}HINT1(51)--HINT1(55): the four momentum and mass ($p_x,p_y,p_z,E,M$)
                (GeV/$c$, GeV, GeV/$c^2$) of the second scattered parton
                in the latest hard scattering of the latest event.
\item{}HINT1(56): $P_T$ (GeV/$c$) of the second scattered parton in the
                latest hard scattering of the latest event.
\item{}HINT1(57)--HINT1(58): not used.
\item{}HINT1(59): (mb) the averaged cross section of the
                triggered jet production (with $P_T$ specified by HIPR1(10)
                and with switch by IHPR2(3)) per nucleon-nucleon
                collision,
                $\int d^2b\{1-\exp[-\sigma_{jet}^{trig}T_N(b)]\}$
\item{}HINT1(60): (mb) the averaged inclusive cross section of the
                triggered jet production $\sigma_{jet}^{trig}$
                (with $P_T$ specified by
                HIPR1(10) and with switch by IHPR2(3)) per
                nucleon-nucleon collision.
\item{}HINT1(61): (mb) the triggered jet production cross section without
                nuclear shadowing effect (similar to HINT1(14)).
\item{}HINT1(62): (mb) the cross section to account for the projectile
                shadowing correction term in the triggered jet cross
                section (similar to HINT1(15)).
\item{}HINT1(63): (mb) the cross section to account for the target
                shadowing correction term in the triggered jet cross
                section (similar to HINT1(16)).
\item{}HINT1(64): (mb) the cross section to account for the cross
                term of shadowing correction in the triggered jet
                cross section (similar to HINT1(17).
\item{}HINT1(65): (mb) the inclusive cross section for latest triggered
                jet production which depends on the transverse coordinates
                of the colliding nucleons (similar to HINT1(18)).
\item{}HINT1(67)--HINT1(71): not used.
\item{}HINT1(72)--HINT1(75): three parameters for the Wood-Saxon
                projectile nuclear distribution and the normalization
                read from a table inside the program,
                $\rho(r)=C[1+W(r/R_A)^2]/\{1+\exp[(r-R_A)/D]\}$,
                $R_A$=HINT1(72), $D$=HINT1(73), $W$=HINT1(74), $C$=HINT1(75).
\item{}HINT1(76)--HINT1(79): three parameters for the Wood-Saxon
                projectile nuclear distribution and the normalization
                read from a table inside the program,
                $\rho(r)=C[1+W(r/R_A)^2]/\{1+\exp[(r-R_A)/D]\}$,
                $R_A$=HINT1(76), $D$=HINT1(77), $W$=HINT1(78), $C$=HINT1(79).
\item{}HINT1(80)--HINT1(100): the probability of $j=0-20$ number of hard
                scatterings per nucleon-nucleon collisions.
\item{}IHNT2(1): the mass number of the projectile nucleus (1 for a hadron).
\item{}IHNT2(2): the charge number of the projectile nucleus. If the
                projectile is a hadron, it gives the charge of the hadron.
\item{}IHNT2(3): the mass number of the target nucleus (1 for a hadron).
\item{}IHNT2(4): the charge number of the target nucleus. If the target
                is a hadron, it gives the charge of the hadron.
\item{}IHNT2(5): the flavor code of the projectile hadron (0 for nucleus).
\item{}IHNT2(6): the flavor code of the target hadron (0 for nucleus).
\item{}IHNT2(7)--IHNT2(8): not used.
\item{}IHNT2(9): the flavor code of the first scattered parton in the
                triggered hard scattering.
\item{}IHNT2(10): the flavor code of the second scattered parton in the
                triggered hard scattering.
\item{}IHNT2(11): the sequence number of the projectile nucleon in the
                latest nucleon-nucleon interaction of the latest event.
\item{}IHNT2(12): the sequence number of the target nucleon in the latest
                nucleon-nucleon interaction of the latest event.
\item{}IHNT2(13): status of the latest soft string excitation.
        \vspace{-12.0pt}
        \begin{description}
        \itemsep=-4.0pt
                \item{}=1: double diffractive.
                \item{}=2: single diffractive.
                \item{}=3: non-single diffractive.
        \end{description}
        \vspace{-4.0pt}
\item{}IHNT2(14): the flavor code of the first scattered parton in the
                latest hard scattering of the latest event.
\item{}IHNT2(15): the flavor code of the second scattered parton in the
                latest hard scattering of the latest event.
\item{}IHNT2(16)--IHNT2(50): not used.

\end{description}

\subsection{Other physics routines}

        Inside HIJING main routines, calls have to be made to many other
routines to carry out the specified simulations. We give here a brief
description of some of those routines.

\begin{description}
\itemsep=-4.0pt
\item{}SUBROUTINE HIJINI
\item{}Purpose: to reset all relevant common blocks and variables and
        initialize the program for each event.
\end{description}

\begin{description}
\itemsep=-4.0pt
\item{}SUBROUTINE HIJCRS
\item{}Purpose: to calculate cross sections of minijet production,
        cross section of the triggered processes,
        elastic, inelastic and total cross section of nucleon-nucleon
        (or hadron) collisions within the eikonal formalism.
\end{description}

\begin{description}
\itemsep=-4.0pt
\item{}SUBROUTINE JETINI (I\_TYPE)
\item{}Purpose: to initialize the program for generating hard scatterings
        as specified by the parameters and options.
\end{description}

\begin{description}
\itemsep=-4.0pt
\item{}SUBROUTINE HIJHRD (JP, JT, JOUT, JFLG, IOPJET0)
\item{}Purpose: to simulate one hard scattering among the multiple jet
        production per nucleon-nucleon (hadron) collision and the
        associated radiations by calling PYTHIA subroutines.
\end{description}

\begin{description}
\itemsep=-4.0pt
\item{}SUBROUTINE HARDJET (JP, JT, JFLG)
\item{}Purpose: to simulate the triggered hard processes.
\end{description}

\begin{description}
\itemsep=-4.0pt
\item{}SUBROUTINE HIJSFT (JP, JT, JOUT, IERROR)
\item{}Purpose: to generate the soft interaction for each binary
        nucleon-nucleon collision.
\end{description}

\begin{description}
\itemsep=-4.0pt
\item{}SUBROUTINE HIJSRT (JPJT, NPT)
\item{}Purpose: to rearrange the gluon jets in a string system according
        to their rapidities.
\end{description}

\begin{description}
\itemsep=-4.0pt
\item{}SUBROUTINE QUENCH (JPJT, NTP)
\item{}Purpose: to perform jet quenching by allowing final state
        interaction of produced jet inside the excited strings.
        The energy lost by the jets will be transferred to other
        string systems.
\end{description}

\begin{description}
\itemsep=-4.0pt
\item{}SUBROUTINE HIJFRG (JTP, NTP, IERROR)
\item{}Purpose: to arrange the produced partons together with the
        valence quarks and diquarks (anti-quarks) and LUEXEC subroutine in
        JETSET is called to perform the fragmentation for each string
        system.
\end{description}

\begin{description}
\itemsep=-4.0pt
\item{}SUBROUTINE ATTRAD (IERROR)
\item{}Purpose: to perform soft radiations according to Lund dipole
        approximation.
\end{description}

\begin{description}
\itemsep=-4.0pt
\item{}SUBROUTINE ATTFLV (ID, IDQ, IDQQ)
\item{}Purpose: to generate the flavor codes of the valence quark and
        diquark (anti-quark) inside a given nucleon (hadron).
\end{description}

\begin{description}
\itemsep=-4.0pt
\item{}SUBROUTINE HIJCSC (JP, JT)
\item{}Purpose: to perform elastic scatterings and possible elastic
        nucleon-nucleon cascading.
\end{description}

\begin{description}
\itemsep=-4.0pt
\item{}SUBROUTINE HIJWDS (IA, IDH, XHIGH)
\item{}Purpose: to set up a distribution function according to the
        three-parameter Wood-Saxon distribution to generate the
        coordinates of the nucleons inside the projectile or
        target nucleus.
\end{description}

\begin{description}
\itemsep=-4.0pt
\item{}FUNCTION  PROFILE (XB)
\item{}Purpose: gives the overlap profile function of two colliding
        nuclei at given impact parameter XB. This can be used to
        weight the simulated events of uniformly distributed impact
        parameter and obtain the results of the minimum biased events.
\end{description}

\begin{description}
\itemsep=-4.0pt
\item{}SUBROUTINE HIBOOST
\item{}Purpose: to transform the produced particles from c.m. frame
        to the laboratory frame.
\end{description}

\begin{description}
\itemsep=-4.0pt
\item{}BLOCK DATA HIDATA
\item{}Purpose: to give the default values of the parameters and
        options and initialize the event record common blocks.
\end{description}

\subsection{Other common blocks}

        There also other two common blocks which contain information
users may find useful.

\begin{description}
\itemsep=-4.0pt
\item{}COMMON/HIJJET4/NDR,IADR(900,2),KFDR(900),PDR(900,5)
\item{}Purpose: contains information about directly produced particles
                (currently only direct photons).
\item{}NDR: total number of directly produced particles.
\item{}IADR(I, 1), IADR(I, 2): the sequence numbers of projectile and
                target nucleons which produce particle I during their
                interaction.
\item{}KFDR(I): the flavor code of directly produced particle I.
\item{}PDR(I, 1,$\cdots$,5): four momentum and mass ($p_x,p_y,p_z,E,M$)
                (GeV/c, GeV, GeV/$c^2$) of  particle I.
\end{description}

\begin{description}
\itemsep=-4.0pt
\item{}COMMON/HIJCRDN/YP(3,300),YT(3,300)
\item{}Purpose: to specify the space coordinates of projectile and
                target nucleons inside their parent nuclei.
\item{}YP(1,$\cdots$,3, I): $x,y,z$ (fm) coordinates of the number
                I projectile nucleon relative to the center of its
                parent nucleus.
\item{}YT(1,$\cdots$,3, I): $x,y,z$ (fm) coordinates of the number I target
                nucleon relative to the center of its parent nucleus.
\end{description}

\section{Instruction on How to Use the Program}

        HIJING program was designed for high energy $pp$, $pA$ and $AB$
collisions. It is relatively easy to use with only two main
subroutines and a few adjustable parameters. In this section we will
give three example programs for generating events of fixed impact
parameter, minimum bias, and triggered hard processes. In all the
cases, users should write their own main program with all the relevant
common blocks included. To study the event, users may  have to call
some routines in JETSET. Therefore, knowledge of JETSET will be helpful.
Two special routines of JETSET which users may frequently use are
function LUCHGE(KF) to give three times the charge, and function
ULMASS(KF) to give the mass for a particle/parton with flavor code
KF.

\subsection{Fixed impact parameter}

        For relativistic hadron-nucleus and heavy ion
collisions, events at fixed impact
parameter especially central collisions with $b=0$ are most commonly
studied. It is also the simplest simulation for HIJING. For $pp$
collisions, one should always use zero impact parameter and HIJING
will give the results averaged over the impact parameter. In the
following example program, we generate 1000 central events of
$Au+Au$ at $\sqrt{s}=200$ GeV/n and calculate the rapidity and
transverse momentum distributions of produced charged particles.
The projectile and target nucleons in the beam directions which
have not suffered any interaction are not considered produced
particles. The output of the event-averaged rapidity and transverse
momentum distributions are plotted in Figs.1 and 2.

{\tt
\begin{tabbing}
AAAAA\=AAA\=  \kill
        \> \>CHARACTER FRAME*8, PROJ*8, TARG*8 \\
\>\>    DIMENSION DNDPT(50),DNDY(50)\\
\>\>    COMMON/HIPARNT/HIPR1(100), IHPR2(50), HINT1(100), IHNT2(50) \\
C....information of produced particles: \> \>\\
        \> \>COMMON/HIMAIN1/NATT, EATT, JATT, NT, NP, N0, N01, N10, N11 \\
        \> \>COMMON/HIMAIN2/KATT(130000,4), PATT(130000,4) \\
C....information of produced partons: \> \> \\
\> \>COMMON/HIJJET1/NPJ(300), KFPJ(300,500), PJPX(300,500), \\
     \>\&  \> PJPY(300,500), PJPZ(300,500), PJPE(300,500), PJPM(300,500),\\
     \>\&  \> NTJ(300), KFTJ(300,500), PJTX(300,500), PJTY(300,500),\\
     \>\&  \> PJTZ(300,500), PJTE(300,500), PJTM(300,500)\\
\> \> COMMON/HIJJET2/NSG, NJSG(900), IASG(900,3), K1SG(900,100),\\
\>\&\>K2SG(900,100), PXSG(900,100), PYSG(900,100), PZSG(900,100), \\
\>\&\>PESG(900,100), PMSG(900,100) \\
\> \> COMMON/HISTRNG/NFP(300,15), PP(300,15), NFT(300,15), PT(300,15) \\
C....initialize HIJING for Au+Au collisions at c.m. energy of 200 GeV: \> \>\\
\>\>    EFRM=200.0 \\
\>\>    FRAME='CMS' \\
\>\>    PROJ='A' \\
\>\>    TARG='A' \\
\>\>    IAP=197 \\
\>\>    IZP=79 \\
\>\>    IAT=197 \\
\>\>    IZT=79 \\
\>\>    CALL HIJSET (EFRM, FRAME, PROJ, TARG, IAP, IZP, IAT, IZT) \\
C....generating 1000 central events:\>\> \\
\>\>    N\_EVENT=1000 \\
\>\>    BMIN=0.0 \\
\>\>    BMAX=0.0 \\
\>\>    DO 2000 J=1,N\_EVENT\\
\>\>    \hspace{24pt}CALL HIJING (FRAME, BMIN, BMAX) \\
C....calculate rapidity and transverse momentum distributions of \> \> \\
C....produced charged particles: \>\> \\
\>\>    \hspace{24pt}DO 1000 I=1,NATT \\
C........\>\>exclude beam nucleons as produced particles: \\
\>\>    \hspace{48pt}IF(KATT(I,2).EQ.0 .OR. KATT(I,2).EQ.10) GO TO 1000 \\
C........\>\>select charged particles only: \\
\>\>    \hspace{48pt}IF (LUCHGE(KATT(I,1)) .EQ. 0) GO TO 1000 \\
\>\>    \hspace{48pt}PTR=SQRT(PATT(I,1)**2+PATT(I,2)**2)\\
\>\>    \hspace{48pt}IF (PTR .GE. 10.0) GO TO 100\\
\>\>    \hspace{48pt}IPT=1+PTR/0.2\\
\>\>    \hspace{48pt}DNDPT(IPT)=DNDPT(IPT)+1.0/FLOAT(N\_EVENT)/0.2/2.0/PTR\\
100\>\> \hspace{48pt}Y=0.5*LOG((PATT(I,4)+PATT(I,3))/(PATT(I,4)+PATT(I,3)))\\
\>\>    \hspace{48pt}IF(ABS(Y) .GE. 10.0) GO GO 1000\\
\>\>    \hspace{48pt}IY=1+ABS(Y)/0.2\\
\>\>    \hspace{48pt}DNDY(IY)=DNDY(IY)+1.0/FLOAT(N\_EVENT)/0.2/2.0\\
1000\>\>\hspace{24pt}CONTINUE \\
2000\>\>CONTINUE \\
C....print out the rapidity and transverse momentum distributions:\>\>\\
\>\>    WRITE(*,*) (0.2*(K-1),DNDPT(K),DNDY(K),K=1,50)\\
\>\>    STOP \\
\>\>    END
\end{tabbing}
                }

\subsection{Minimum bias events}

        Because of the diffused distribution of large nuclei, minimum
bias events are dominated by those of large impact parameters with a
long shoulder for small impact parameter events.
To effectively study minimum
bias events, one can generate events uniformly between zero and
the largest impact parameter $R_A+R_B$, and then weight the events by
a Glauber probability,
\begin{equation}
        \frac{1}{\sigma_{AB}}d^2b\{1-\exp[-\sigma_{in}T_{AB}(b)]\}
\end{equation}
where $\sigma_{in}$ is the inelastic cross section for $N$-$N$ collisions and
$\sigma_{AB}$ is the total reaction cross section for $AB$ collisions
integrated over all impact parameters. To obtain the Glauber distribution
a routine named FUNCTION PROFILE(XB) has to be called.

        In the following main program, a range of impact parameters
from 0 to $2R_A$ is divided into 100 intervals. For each fixed
impact parameter, 10 events are generated for $Au+Au$ at $\sqrt{s}=200$ GeV/n.
Then $P_T$ distribution for charged pions is calculated for
the minimum bias events.

{\tt
\begin{tabbing}
AAAAA\=AAA\=  \kill
        \> \>CHARACTER FRAME*8, PROJ*8, TARG*8 \\
        \> \>COMMON/HIPARNT/HIPR1(100), IHPR2(50), HINT1(100), IHNT2(50) \\
        \> \>COMMON/HIMAIN1/NATT, EATT, JATT, NT, NP, N0, N01, N10, N11 \\
        \> \>COMMON/HIMAIN2/KATT(130000,4), PATT(130000,4) \\
        \> \>DIMENSION GB(101), XB(101), DNDP(50) \\
C....initialize HIJING for Au+Au collisions at c.m. energy of 200 GeV: \> \>\\
\>\>    EFRM=200.0 \\
\>\>    FRAME='CMS' \\
\>\>    PROJ='A' \\
\>\>    TARG='A' \\
\>\>    IAP=197 \\
\>\>    IZP=79 \\
\>\>    IAT=197 \\
\>\>    IZT=79 \\
\>\>    CALL HIJSET (EFRM, FRAME, PROJ, TARG, IAP, IZP, IAT, IZT) \\
C....set BMIN=0 and BMAX=R\_A+R\_B \>\> \\
\>\>    BMIN=0.0 \\
\>\>    BMAX=HIPR1(34)+HIPR1(35) \\
C....calculate the Glauber probability and its integrated value:\>\> \\
\>\>    DIP=(BMAX-BMIN)/100.0 \\
\>\>    GBTOT=0.0 \\
\>\>    DO 100 I=1,101 \\
\>\>    \hspace{24pt}XB(I)=BMIN+(I-1)*DIP \\
\>\>    \hspace{24pt}OV=PROFILE(XB(I)) \\
\>\>    \hspace{24pt}GB(I)=XB(I)*(1.0-EXP(-HINT1(12)*OV)) \\
\>\>    \hspace{24pt}GBTOT=GBTOT+GB(I) \\
100\>\> CONTINUE \\
C....generating 10 events for each of 100 impact parameter intervals:\>\> \\
\>\>    NONT=0 \\
\>\>    GNORM=GBTOT \\
\>\>    N\_EVENT=10 \\
\>\>    DO 300 IB=1,100 \\
\>\>    \hspace{24pt}B1=XB(IB) \\
\>\>    \hspace{24pt}B2=XB(IB+1)\\
C.......\>\>normalized Glauber probability:\\
\>\>    \hspace{24pt}W\_GB=(GB(IB)+GB(IB+1))/2.0/GBTOT\\
\>\>    \hspace{24pt}DO 200 IE=1,N\_EVENT\\
\>\>    \hspace{48pt}CALL HIJING(FRAME,B1,B2) \\
C........\>\>count number of events without any interaction\\
C........\>\>and renormalize the total Glauber probability:\\
\>\>    \hspace{48pt}IF (NP+NT .EQ. 0) THEN \\
\>\>    \hspace{62pt}NONT=NONT+1 \\
\>\>    \hspace{62pt}GNORM=GNORM-GB(IB)/FLOAT(N\_EVENT) \\
\>\>    \hspace{62pt}GO TO 200\\
\>\>    \hspace{48pt}ENDIF \\
C....calculate pt distribution of charged pions: \>\> \\
\>\>    \hspace{48pt}DO 150 K=1,NATT \\
C........\>\>select charged pions only: \\
\>\>    \hspace{62pt}IF (ABS(KATT(K,1)) .NE. 211) GO TO 150 \\
C........\>\>calculate pt: \\
\>\>    \hspace{62pt}PTR=SQRT(PATT(K,1)**2+PATT(K,2)**2) \\
C........\>\>calculate pt distribution and weight with normalized\\
C........\>\>Glauber probability to get minimum bias result:\\
\>\>    \hspace{62pt}IF (PTR .GE. 10.0) GO TO 150 \\
\>\>    \hspace{62pt}IPT=1+PTR/0.2 \\
\>\>    \hspace{62pt}DNDP(IPT)=DNDP(IPT)+1.0/W\_GB/FLOAT(N\_EVENT)/0.2 \\
150\>\> \hspace{48pt}CONTINUE \\
200\>\> \hspace{24pt}CONTINUE \\
300\>\> CONTINUE \\
C....renormalize the distribution by the renormalized Glauber \>\> \\
C....probability which excludes the events without any interaction: \>\> \\
\>\>    IF(NONT.NE.0) THEN \\
\>\>    \hspace{24pt}DO 400 I=1,50 \\
\>\>    \hspace{48pt}DNDP(I)=DNDP(I)*GBTOT/GNORM \\
400\>\> \hspace{24pt}CONTINUE \\
\>\>    ENDIF \\
\>\>    STOP \\
\>\>    END
\end{tabbing}
                }

\subsection{Events with triggered hard processes}

        Sometimes, users may want to study events with a hard process.
Since these processes, especially with large transverse momentum, have
very small cross section, it is very inefficient to sort them out among
huge number of ordinary events. However, in HIJING, one can trigger
on such events and generate one hard process in each event with the
background correctly incorporated. One can then calculate the absolute
cross section of such events by using the information stored in
HINT(12) (inelastic $N$-$N$ cross section) and HINT1(59) ( cross section
of triggered process in $N$-$N$ collisions). HIPR1(10) is used to
specify the $P_T$ value or its range.

        In the current version, both large $P_T$ jets (IHPR2(3)=1)
and direct photon production (IHPR2(3)=2) are included. In the
following, we give an example on how to generate a pair of large $P_T$
jets above 20 GeV/$c$ in a central $Au+Au$ collision
at $\sqrt{s}=200$ GeV/n.

{\tt
\begin{tabbing}
AAAAA\=AAA\=  \kill
        \> \>CHARACTER FRAME*8, PROJ*8, TARG*8 \\
\>\>    COMMON/HIPARNT/HIPR1(100), IHPR2(50), HINT1(100), IHNT2(50) \\
C.....switch off jet quenching: \>\> \\
\>\>    IHPR2(4)=0 \\
C.....switch on triggered jet production: \>\>\\
\>\>    IHPR2(3)=1 \\
C.....set the pt range of the triggered jets: \>\> \\
\>\>    HIPR1(10)=-20 \\
C....initialize HIJING for Au+Au collisions at c.m. energy of 200 GeV: \> \>\\
\>\>    EFRM=200.0 \\
\>\>    FRAME='CMS' \\
\>\>    PROJ='A' \\
\>\>    TARG='A' \\
\>\>    IAP=197 \\
\>\>    IZP=79 \\
\>\>    IAT=197 \\
\>\>    IZT=79 \\
\>\>    CALL HIJSET (EFRM, FRAME, PROJ, TARG, IAP, IZP, IAT, IZT) \\
C....generating one central event with triggered jet production:\>\> \\
\>\>    BMIN=0.0 \\
\>\>    BMAX=0.0 \\
\>\>    CALL HIJING (FRAME, BMIN, BMAX) \\
C....print out flavor code of the first jet:\>\> \\
\>\>    WRITE(*,*) IHNT2(9) \\
C....and its four momentum:\>\> \\
\>\>    WRITE(*,*) HINT1(21), HINT1(22), HINT1(23), HINT1(24) \\
C....print out flavor code of the second jet:\>\> \\
\>\>    WRITE(*,*) IHNT2(10) \\
C....and its four momentum:\>\> \\
\>\>    WRITE(*,*) HINT1(31), HINT1(32), HINT1(33), HINT1(34) \\
\>\>    STOP \\
\>\>    END
\end{tabbing}
                }

\section*{Acknowledgements}

        During the development of this program, we benefited a lot
from discussions with J.~Carroll, J.~W.~Harris, P.~Jacobs,
M.~A.~Bloomer, and A.~Poskanzer.
We would like to thank T.~Sj\"{o}strand for making available
JETSET and PYTHIA Monte Carlo programs on which HIJING is based on.
We would also like to thank K.~J.~Eskola for helpful comments and
discussions.

\section*{Appendix: Flavor Code}

        For users' reference, a selection of flavor codes from JETSET 7.2
are listed below. For full list please check JETSET documentation.
The codes for anti-particles are just the negative values of the
corresponding particles.

\begin{tabbing}
bbbbbbbbbbbbbbb\=bbbbbb\=bbbbbbbbbbbbbb\=bbbbbb\= \kill
Quarks and leptons \> \> \> \> \\
\>              1       \>d             \>11    \>$e^-$ \\
\>              2       \>u             \>12    \>$\nu_e$ \\
\>              3       \>s             \>13    \>$\mu^-$ \\
\>              4       \>c             \>14    \>$\nu_{\mu}$ \\
\>              5       \>b             \>15    \>$\tau^-$ \\
\>              6       \>t             \>16    \>$\nu_{\tau}$ \\
\>\>\>\> \\
Gauge bosons \>\>\>\> \\
\>              21      \>g \>\> \\
\>              22      \>$\gamma$ \>\> \\
\>\>\>\> \\
Diquarks \>\>\>\> \\
\>                      \>              \>1103  \>dd$_1$ \\
\>              2101    \>ud$_0$        \>2103  \>ud$_1$ \\
\>                      \>              \>2203  \>uu$_1$ \\
\>              3101    \>sd$_0$        \>3103  \>sd$_1$ \\
\>              3201    \>su$_0$        \>3203  \>su$_1$ \\
\>                      \>              \>3303  \>ss$_1$ \\
\>\>\>\> \\
Mesons \>\>\>\> \\
\>              211     \>$\pi^+$       \>213   \>$\rho^+$ \\
\>              311     \>K$^0$         \>313   \>K$^{*0}$ \\
\>              321     \>K$^+$         \>323   \>K$^{*+}$ \\
\>              411     \>D$^+$         \>413   \>D$^{*+}$ \\
\>              421     \>D$^0$         \>423   \>D$^{*0}$ \\
\>              431     \>D$_{\mbox{s}}^+$
                                \>433   \>D$_{\mbox{s}}^{*+}$ \\
\>              511     \>B$^0$         \>513   \>B$^{*0}$ \\
\>              521     \>B$^+$         \>523   \>B$^{*+}$ \\
\>              531     \>B$_{\mbox{s}}^0$
                                \>533   \>B$_{\mbox{s}}^{*0}$ \\
\>              111     \>$\pi^0$       \>113   \>$\rho^0$ \\
\>              221     \>$\eta$        \>223   \>$\omega$ \\
\>              331     \>$\eta'$       \>333   \>$\phi$ \\
\>              441     \>$\eta_{\mbox{c}}$     \>443   \>J/$\psi$ \\
\>              551     \>$\eta_{\mbox{b}}$     \>553   \>$\Upsilon$ \\
\>              661     \>$\eta_{\mbox{t}}$     \>663   \>$\Theta$ \\
\>              130     \>K$_L^0$ \>\> \\
\>              310     \>K$_S^0$ \>\> \\
\>\>\>\> \\
Baryons \>\>\>\> \\
\>                      \>              \>1114  \>$\Delta^-$ \\
\>              2112    \>n             \>2114  \>$\Delta^0$ \\
\>              2212    \>p             \>2214  \>$\Delta^+$ \\
\>                      \>              \>2224  \>$\Delta^{++}$ \\
\>              3112    \>$\Sigma^-$    \>3114  \>$\Sigma^{*-}$ \\
\>              3122    \>$\Lambda^0$   \> \> \\
\>              3212    \>$\Sigma^0$    \>3214  \>$\Sigma^{*0}$ \\
\>              3222    \>$\Sigma^+$    \>3224  \>$\Sigma^{*+}$ \\
\>              3312    \>$\Xi^-$       \>3314  \>$\Xi^{*-}$ \\
\>              3322    \>$\Xi^0$       \>3324  \>$\Xi^{*0}$ \\
\>                      \>              \>3334  \>$\Omega^-$ \\
\>              4112    \>$\Sigma_{\mbox{c}}^0$
                                \>4114  \>$\Sigma_{\mbox{c}}^{*0}$ \\
\>              4122    \>$\Lambda_{\mbox{c}}^+$ \> \> \\
\>              4212    \>$\Sigma_{\mbox{c}}^+$
                                \>4214  \>$\Sigma_{\mbox{c}}^{*+}$ \\
\>              4222    \>$\Sigma_{\mbox{c}}^{++}$
                                \>4224  \>$\Sigma_{\mbox{c}}^{*++}$ \\
\>              4132    \>$\Xi_{\mbox{c}}^0$    \> \> \\
\>              4312    \>$\Xi'$$_{\mbox{c}}^0$
                                \>4314  \>$\Xi_{\mbox{c}}^{*0}$ \\
\>              4232    \>$\Xi_{\mbox{c}}^+$    \> \> \\
\>              4322    \>$\Xi'$$_{\mbox{c}}^+$
                                \>4324  \>$\Xi_{\mbox{c}}^{*+}$ \\
\>              4332    \>$\Omega_{\mbox{c}}^0$
                                \>4334  \>$\Omega_{\mbox{c}}^{*0}$ \\
\>              5112    \>$\Sigma_{\mbox{b}}^-$
                                \>5114  \>$\Sigma_{\mbox{b}}^{*-}$ \\
\>              5122    \>$\Lambda_{\mbox{b}}^0$        \>\> \\
\>              5212    \>$\Sigma_{\mbox{b}}^0$
                                \>5214  \>$\Sigma_{\mbox{b}}^{*0}$ \\
\>              5222    \>$\Sigma_{\mbox{b}}^+$
                                \>5224  \>$\Sigma_{\mbox{b}}^{*+}$
\end{tabbing}

\pagebreak

\pagebreak

{\noindent\Large\bf Figure Captions}
\vspace{24pt}
\begin{description}

\item[Fig. 1] The rapidity distribution of charged particles produced in
              central $Au+Au$ collisions at $\sqrt{s}=200$ GeV/n, obtained
              from the example program for fixed impact parameter.

\item[Fig. 2] The transverse momentum distribution of charged particles
              in central $Au+Au$ collisions, obtained from the example
              program for fixed impact parameter.

\end{description}

\end{document}